# Item Association Factorization Mixed Markov Chains for Sequential Recommendation

DongYu Du, Yue Chan*

**Abstract**—Sequential recommendation refers to recommending the next item of interest for a specific user based on his/her historical behavior sequence up to a certain time. While previous research has extensively examined Markov chain-based sequential recommendation models, the majority of these studies have focused on the user's historical behavior sequence but have paid little attention to the overall correlation between items. This study introduces a sequential recommendation algorithm known as Item Association Factorization Mixed Markov Chains, which incorporates association information between items using an item association graph, integrating it with user behavior sequence information. Our experimental findings from the four public datasets demonstrate that the newly introduced algorithm significantly enhances the recommendation ranking results without substantially increasing the parameter count. Additionally, research on tuning the prior balancing parameters underscores the significance of incorporating item association information across different datasets.

* unimelbat@hotmail.com

## 1 INTRODUCTION

As consumer products continue to surge, users face increasing difficulty in identifying products of their genuine interest from a wide range of commodities. Consequently, commercial platforms have widely implemented recommendation algorithms to provide users personalized product lists [1]. In the early stages of such research, recommendation algorithms primarily utilized the collaborative filtering approach, which focused on the static relationships between users and items, ignoring temporal changes in user preferences. Later on, some algorithms began to take into account sequential patterns in historical data to predict the likelihood of the next item, namely sequential recommendation algorithms [2], which have attracted significant attention from the academic community due to their exceptional performance. These algorithms can be classified as recommendation methods based on Collaborative Filtering [3, 4], Markov Chains models [5–7], translation-based models [8, 9], content-aware sequential models [10–12], and deep learning-based models [13–17].

This paper conducts an in-depth exploration of Markov Chains and proposes optimizations to Fossil [7]. Fossil integrates both FISM [18] and Markov Chains, wherein FISM is employed to capture long-term interests, whereas Markov Chains are utilized to model short-term interests. While Fossil has demonstrated strong performance across various datasets, such method has limitations in terms of exploring the item associations. In other words, it falls short in capturing the nuanced similarities between different items. Accordingly, this paper introduces Item Association Factorization Mixed Markov Chains for sequential recommendation (or IAFMC), with the addition of minimal training parameters, factorizes the association information between items so as to integrate it into item sequence representations. This method significantly improves the prediction accuracy of next-item recommendations by effectively incorporating item associations, thereby addressing challenges associated with sparse data and user cold start.

In the second section, this paper will provide an overview of various recommendation algorithms and their associated theories. The third section will elaborate on the model and algorithm proposed in this paper, followed by the extensive experimental validation on four public datasets in the fourth section. Our numerical outcomes indicate that IAFMC outperforms a range of comparable models across most metrics. We also conducted a study on balancing parameters to verify the significance of the newly introduced association information between items, based on different datasets. Finally, we will summarize and analyze IAFMC in the concluding section.

## 2 RELATED WORK

### 2.1 Implicit Feedback

Different algorithms utilize various data information for modeling [19]. Some algorithms leverage explicit user preference information, such as user ratings [20], to model such preferences. These algorithms can be broadly categorized into user-oriented [21–23] and item-oriented [24, 25] approaches. On the other hand, some algorithms focus on the utilization of simple and easily accessible user information, such as historical user behavior data like clicks, views, and so on [26]. This type of data is referred to as implicit feedback, as it typically does not explicitly convey user preferences, which is often represented as binary tuples (user, item). In such cases, for a particular user, we consider all the items that the user has interacted with to be equal and positive.

Matrix factorization-based methods are often used to solve the recommendation problem with implicit feedback. Hu et al. [27] and Pan et al. [28] proposed a matrix factorization method with case weights, but this method did not directly optimize for personalized ranking tasks. Rendle et al. [29] proposed the use of paired loss to optimize the model by comparing the different recommendation priorities of two items, maximizing the prediction gap between positive and negative feedback. The aforementioned algorithms only explore the linear features in the data. Last but not least, there are some deep learning-based methods, such as NeuMF [30] or LightGCN [31], that attempt to explore the potential nonlinear features in the data.

### 2.2 Sequential Recommendation

People's interests and hobbies evolve gradually over time. In contrast to static recommendation models, sequential recommendation models can better capture the current interests of users so as to recommend suitable items accordingly. Intrinsically, the order in which users click or browse items contains hidden information about the changes in user preferences, so that datasets can be further enriched by using triplets (user id, item id, time) to characterize the historical behavior sequences of users.

Regarding the temporal consideration, some algorithms [32] use explicit timestamps to build models that understand past behavior based on a specific time of a user's preference. Some algorithms do not directly use specific timestamps, but model the sequence of actions directly.

In the early stages of such research, people often used sequential pattern mining to model [33, 34], but it required a lot of computational resources and the design of sequence rules, made it difficult to simulate complex sequence relationships. Koren [35] introduced time factors into the factor model based on collaborative filtering, explicitly modeling changes in user interests, and significantly improving performance compared to traditional methods. When it comes to modeling time series, we can think of the Markov chain model. It is often used to solve sequential prediction tasks, revealing sequence patterns [36], and directly modeling decision processes [37].

With regard to sequence recommendation, user interaction sequences are often treated as sentences, and the probability of a sequence can be calculated using Bayesian formulas. Rendle et al. [5] proposed FPMC to decompose the Markov chain modeling and combined it with traditional matrix factorization for a general representation of users and items. He et al. [7] also proposed Fossil, which integrates similarity models into sequential recommendation. Fossil further refines parameters sensitive to users and sequences to handle high-order Markov chains. In addition to traditional statistical methods, some researchers leverage mature sequence modeling using deep neural networks for sequential recommendation. For instance, GRU4Rec [14] proposes to use Recurrent Neural Network in enriching session-based recommendation. For more descriptions on various machine learning methods, we refer our readers to Sec. 4.4 for more details.

## 3 PRELIMINARIES AND DISCUSSIONS

In this chapter, we introduce a sequential recommendation model called Item

Association Factorization Mixed Chains for Sequential Recommendation, namely FIAMC. We extract items' association information from the item association graph and integrate such information into a factorized Markov chain model.

## 3.1 Problem Formulation

In one-class collaborative filtering, these actions are considered as positive feedback, such as browsing and clicking. Given n users and m items, each user is associated with a corresponding sequence of actions $S_u = \left\{i_u^1, i_u^2, \ldots, i_u^t, \ldots, i_u^{|S_u|}\right\}$, where $i_u^t$ denotes the t-th item that interacts with user $u$. We further take takes the user's sequence of actions($u$, $S_u$) as input and outputs a ranked list of the top n recommended items for that user $P_u = \{j_u^1, j_u^2, \ldots, j_u^t, \ldots, j_u^n | j_u^t \in I \backslash I_u\}$, where $j_u^t$ represents the t th recommended item. $P_u$ should ideally include the user's next interaction item and be ranked as high as possible. Please refer to the TABLE 1 for an explanation of the symbols used in this paper and their meanings.

TABLE 1: for an explanation of the symbols used in this paper and their meanings.

| Notation | Explanation |
|---|---|
| $n$ | number of users |
| $m$ | number of items |
| $u$ | user ID, $u \in \{1, 2, ..., n\}$ |
| $i$ | item ID, $i \in \{1, 2, ..., n\}$ |
| $\mathcal{U}$ | The whole set of users |
| $\mathcal{I}$ | The whole set of items |
| $\mathcal{P}$ | the whole set of observed $(u, i)$ pairs |
| $\mathcal{A}$ | a sampled set of unobserved $(u, i)$ pairs |
| $\mathcal{I}_u$ | a set of items that have been interacted by user $u$ |
| $\mathcal{I}_u^{te}$ | a set of preferred items by user u in test data |
| $\mathcal{U}^{te}$ | a set of users in test data |
| $\mathcal{M}_i$ | a set of items that associate item $i$ |
| $\mathcal{S}_u$ | $\mathcal{S}_u = \{i_u^1, i_u^2, ..., i_u^t, ..., i_u^{|\mathcal{S}_u|}\}$ |
| $i_u^t$ | the $t$th item in $\mathcal{S}_u$ |
| $\hat{r}_{ui_u^t}$ | predicted preference of user $u$ to item $i_u^t$ |
| $L$ | the order of Markov chains |
| $\ell$ | the $\ell$th order of Markov chains, $\ell \in \{1, 2, ..., L\}$ |
| $i_u^{t-\ell}$ | the $(t-\ell)$th item in $\mathcal{S}_u$ |
| $\eta \in \mathbb{R}^{1 \times L}$ | global weighting vector |
| $\eta^u \in \mathbb{R}^{1 \times L}$ | personalized weighting vector w.r.t. user $u$ |
| $\zeta \in \mathbb{R}$ | global weighting vector |
| $\zeta^i \in \mathbb{R}$ | personalized weighting vector w.r.t. item $i$ |
| $\delta_{ii'}$ | the associated wegiht between item $i$ and item $i' \in \mathcal{M}_i$ |
| $d \in \mathbb{R}$ | number of latent dimensions |
| $T_i, W_i \in \mathbb{R}^{1 \times d}$ | item-specific latent feature vector w.r.t. item $i$ |
| $b_i \in \mathbb{R}$ | item bias |
| $\hat{r}_{ui}$ | predicted preference of user $u$ to item $i$ |
| $\lambda$ | learning rate |
| $\alpha_t, \alpha_w, \beta_b, \beta_\eta, \beta_\zeta$ | tradeoff parameters of regularization terms |
| $T$ | iteration number |
| $\alpha$ | hyperparameter of users |
| $\beta$ | hyperparameter of items |

### 3.2 Basic Method: Fossil

FISM [18] utilizes the formation of items, interacting with users to represent the long-term attributes of users without considering the time series information. It updates the model through the multiplication of two low-dimensional matrices. Building upon FISM, Fossil [7] takes into account a Markov chain to better characterize the changes in the short-term preferences of users, where the fundamental formula is as follows:

$$\hat{r}_{ui_u^t} = b_{i_u^t} + U_{ui_u^t} V_{i_u^t}^T, \qquad (1)$$

where:

$$U_{ui_u^t} = \frac{1}{|\mathcal{I}_u \backslash \{i_u^t\}|^\alpha} \sum_{i' \in \mathcal{I}_u \backslash \{i_u^t\}} W_{i'} + \sum_{\ell=1}^{L} (\eta_\ell + \eta_\ell^u) W_{i_u^t}, \quad (2)$$

where $\eta_\ell^u$ controls the preference weight of user u under different orders, while $\eta_\ell$

represents the global weight. In Fossil, the parameter $\alpha$ is typically set to 0.2.

### 3.3 Our Objective Function

The significant contribution of Fossil lies in contemplating short-term sequence information through higher-order Markov chain. However, it lacks items association. In our work, we transform the sequence information of user interactions with items into associative information for items.

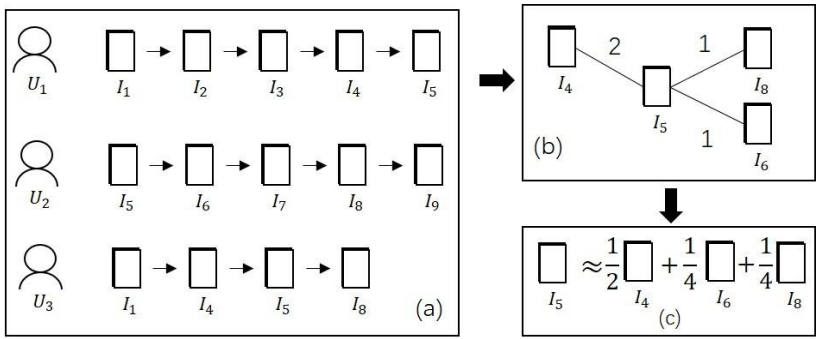

Fig. 1: An example of the transformation of item association information. (a) Show the sequence information of three users; (b) Extract the association information for item $I_5$ from panel (a); (c) Get a new representation of item $I_5$ from panel (b).

As depicted in Fig. 1, three users' (denoted as $U_1$, $U_2$ and $U_3$) sequential interactions with items are shown in panel (a). In the case of item $I_5$, its adjacent items can be derived, as indicated in panel (b), with the frequency of adjacency denoted as the weight. To represent a particular item, its association with neighboring items can be employed, as shown in panel (c) where the item $I_5$ showcases the representation through item association information.

Based on Fossil, we introduce item association information to enhance the effectiveness of recommendations, where item association information is defined as follows:

$$V_{i_u^t} = \frac{1}{|\sum_{i' \in \mathcal{M}_{i_u^t}} \delta_{i_u^t i'}|^\beta} \sum_{i' \in \mathcal{M}_{i_u^t}} \delta_{i_u^t i'} T_{i'} + (\zeta + \zeta^{i_u^t}) T_{i_u^t}, \quad (3)$$

where $\frac{1}{|\sum_{i' \in \mathcal{M}_{i_u^t}} \delta_{i_u^t i'}|^\beta} \sum_{i' \in \mathcal{M}_{i_u^t}} \delta_{i_u^t i'} T_{i'}$ means the item association information as shown in Fig. 1. In the dataset of this paper, we choose to include an additional term $(\zeta + \zeta^{i_u^t}) T_{i_u^t}$ with the aim of emphasizing the intrinsic properties of this item, due to the characteristic that item $i_u^t$, which is not included in $M_{i_u^t}$.

Our primary goal is to rank the participating projects. Based on the pairwise preference assumption [29], when we have $\hat{r}_{ui} > \hat{r}_{uj}$, it implies that user $u$ prefers item $i \in I_u$ over item $j \in I \setminus I_u$ Given that, we can adopt the personalized pairwise ranking to minimize the loss, leading to the objective function:

$$\min_{\Theta} \sum_{u \in \mathcal{U}} \sum_{i_u^t \in \mathcal{S}_u, t \neq 1} \sum_{j \notin \mathcal{I}_u} f_{ui_u^t j}, \qquad (4)$$

$$f_{ui_u^t j} = -\ln \sigma(\hat{r}_{ui_u^t} - \hat{r}_{uj}) + R_{ui_u^t j}, \qquad (5)$$

$$\sigma(x) = \frac{1}{1 + e^{-x}}, \qquad (6)$$

$$R_{ui_u^t j} = \frac{\alpha_t}{2} \sum_{i' \in \mathcal{M}_{i_u^t}} \|T_{i'}\|^2 + \frac{\alpha_t}{2}\|T_{i_u^t}\|^2 + \frac{\alpha_t}{2} \sum_{j' \in \mathcal{M}_j} \|T_{j'}\|^2$$
$$+ \frac{\alpha_t}{2}\|T_j\|^2 + \frac{\alpha_w}{2} \sum_{i' \in \mathcal{I}_u} \|W_{i'}\|^2 + \frac{\alpha_w}{2} \sum_{\ell=1}^{L} \|W_{i_u^{t-\ell}}\|^2$$
$$+ \frac{\beta_b}{2} b_{i_u^t}^2 + \frac{\beta_b}{2} b_j^2 + \frac{\beta_\eta}{2}\|\eta_\ell\|^2 + \frac{\beta_\eta}{2} \sum_{\ell=1}^{L} \|\eta_\ell^u\|^2$$
$$+ \frac{\beta_\zeta}{2}\|\zeta\|^2 + \frac{\beta_\zeta}{2}\|\zeta^{i_u^t}\|^2 + \frac{\beta_\zeta}{2}\|\zeta^j\|^2, \qquad (7)$$

where $-\ln\sigma(\hat{r}_{ui_u^t} - \hat{r}_{uj})$ implies that we aim to maximize the score difference between $\hat{r}_{ui_u^t}$ and $\hat{r}_{uj}$ as much as possible. represents the regularization term designed to alleviate the issue of overfitting.

For $f_{ui_u^t j}$, we obtain the binary tuple $(u, i_u^t)$ by initially randomly sampling a user u and its corresponding item $i_u^t$. We then sample an item j from set $j \in I \backslash I_u$, representing items that have not been interacted with. These components are combined to form the ternary tuple f. Here, the parameter $\Theta = \{T_i, W_i, b_i, \eta_\iota, \eta_\iota^u, \zeta, \zeta^i, i = 1,2,\dots,m; u = 1,2,\dots,n; \iota = 1,2,\dots,L\}$.

### 3.4 Gradients and Update Rules

We will employ the stochastic gradient descent (SGD) method [38] to minimize the objective function in Eq. (4). For each parameter $\theta \in \Theta$, we have $\nabla \theta = \frac{\partial f_{ui_u^t j}}{\partial \theta}$, and the specific calculation is shown as follows:

$$V_j = \frac{1}{|\sum_{i' \in \mathcal{M}_j} \delta_{ji'}|^\beta} \sum_{i' \in \mathcal{M}_j} \delta_{ji'} T_{i'} + (\zeta + \zeta^j) T_j. \quad (8)$$

$$\nabla b_{i_u^t} = \beta_v b_{i_u^t} + (-1)\sigma(\hat{r}_{uj} - \hat{r}_{ui_u^t}). \quad (9)$$

$$\nabla b_j = \beta_v b_j + \sigma(\hat{r}_{uj} - \hat{r}_{ui_u^t}). \quad (10)$$

$$\nabla W_{i'} = \alpha_w W_{i'} + (-1)\sigma(\hat{r}_{uj} - \hat{r}_{ui_u^t}) \\ (\frac{1}{\sqrt{|\mathcal{I}_u \setminus \{i_u^t\}|}} V_{i_u^t} - \frac{1}{\sqrt{|\mathcal{I}_u|}} V_j). \quad (11)$$

$$\nabla W_{i_u^{t-\ell}} = \alpha_w W_{i_u^{t-\ell}} + (-1)\sigma(\hat{r}_{uj} - \hat{r}_{ui_u^t}) \\ \sum_{\ell=1}^{L} (\eta_\ell + \eta_\ell^u)(V_{i_u^t} - V_j). \quad (12)$$

$$\nabla \eta_\ell = \beta_\eta \eta_\ell + (-1)\sigma(\hat{r}_{uj} - \hat{r}_{ui_u^t}) \sum_{\ell=1}^{L} W_{i_u^{t-\ell}}(V_{i_u^t} - V_j). \quad (13)$$

$$\nabla \eta_\ell^u = \beta_\eta \eta_\ell^u + (-1)\sigma(\hat{r}_{uj} - \hat{r}_{ui_u^t}) \sum_{\ell=1}^{L} W_{i_u^{t-\ell}}(V_{i_u^t} - V_j). \quad (14)$$

$$\nabla T_{i'} = \alpha_t T_{i'} + (-1)\sigma(\hat{r}_{uj} - \hat{r}_{ui_u^t}) \\ (\frac{\delta_{i_u^t i'}}{|\sum_{i' \in \mathcal{M}_{i_u^t}} \delta_{i_u^t i'}|^\beta} U_{ui_u^t} - \frac{\delta_{ji'}}{|\sum_{i' \in \mathcal{M}_j} \delta_{ji'}|^\beta} U_{uj}). \quad (15)$$

$$\nabla T_{i_u^t} = \alpha_t T_{i_u^t} + (-1)\sigma(\hat{r}_{uj} - \hat{r}_{ui_u^t})(\zeta + \zeta^{i_u^t}) U_{ui_u^t}. \quad (16)$$

$$\nabla T_j = \alpha_t T_j + \sigma(\hat{r}_{uj} - \hat{r}_{ui_u^t})(\zeta + \zeta^j) U_{uj}. \quad (17)$$

$$\nabla \zeta = \beta_\zeta \zeta + (-1)\sigma(\hat{r}_{uj} - \hat{r}_{ui_u^t})(U_{ui_u^t} T_{i_u^t} - U_{uj} T_j). \quad (18)$$

$$\nabla \zeta_{i_u^t} = \beta_\zeta \zeta_{i_u^t} + (-1)\sigma(\hat{r}_{uj} - \hat{r}_{ui_u^t}) U_{ui_u^t} T_{i_u^t}. \quad (19)$$

$$\nabla \zeta_j = \beta_\zeta \zeta_{i_u^t} + \sigma(\hat{r}_{uj} - \hat{r}_{ui_u^t}) U_{uj} T_j. \quad (20)$$

For each sampled ternary tuple $(u, i_u^t, j)$, we obtain the update rule for each parameter

$\theta = \theta - \lambda\theta$, where $\lambda > 0$ implies the learning rate.

### 3.5 Algorithm

Solving the objective function in Eq. (4) using the commonly used SGD algorithm is illustrated in Algorithm 1. This algorithm consists of two nested loops. Wile the outer loop iterates from 1 to t, aiming to repeatedly learn and optimize the model; the inner loop, we first sample positive instances $(u, i_u^t)$, then randomly select negative instances $j \in I\backslash I_u$, and update the model parameters through computation.

---
**Algorithm 1:** The algorithm of IAFMC
---
1  Initialize the model parameters;
2  **for** $t = 1, \ldots, T$ **do**
3      **for** *each* $(u, i_u^t) \in \mathcal{P}$ *in a random order* **do**
4          Randomly pick up an item $j$ from $\mathcal{I}\backslash\mathcal{I}_u$;
5          Calculate gradients according to (8 - 20);
6          Update the model parameters $T_i, W_i, b_i,$
             $\eta_\ell, \eta_\ell^u, \zeta, \zeta^i$;
7      **end**
8  **end**

---

## 4 EXPERIMENTS

### 4.1 Datasets

We will evaluate IAFMC on the prescribed four public datasets which are derived from Amazon review datasets [39, 40] in different domains, such as toys and tools. The characteristic of the Amazon dataset is its high sparsity. We sort interactions for each user in terms of sequences using timestamps for each rating. The most recent interactions will be used as the test set, while the second most recent inter- actions will serve as the validation set. Following [8, 41], we also filter out users with fewer than 5 interactions using a 5-core setting. We consider all interactions as positive feedback. The details of each dataset are presented in Table 2.

TABLE 2: Datasets Statistics.

| Dataset | users | items | interactions | density | avg. interactions per user |
|---|---|---|---|---|---|
| Beauty | 22,363 | 12,101 | 198,502 | 0.05% | 8.3 |
| Office | 4,905 | 2,420 | 53,258 | 0.44% | 10.8 |
| Tools | 16,638 | 10,217 | 134,476 | 0.08% | 8.1 |
| Toys | 19412 | 11924 | 167,597 | 0.07% | 8.6 |

For each user, we rank the predicted scores calculated by function in Eq. (1) and generate the top-N recommendation list in ascending order. We then employ standard top-N ranking evaluation metrices such as Recall@N and NDCG@N. Our primary focus is to determine the values of these metrics when N is et to 5 and 10.

- Recall@ is the ratio of the number of items in the top-N recommendation list that exactly match the user's preferences to the total number of items the user likes.

$$\text{Rec}@N = \sum_{u \in \mathcal{U}^{te}} \frac{\frac{1}{|\mathcal{I}_u^{te}|}\sum_{\ell=1}^{N} \delta(i(\ell) \in \mathcal{I}_u^{te})}{|\mathcal{U}^{te}|}, \quad (21)$$

where $\delta(x)$ is an indicator function. When x is predicted, $\delta(x) = 1$; otherwise $\delta(x) = 0$.

- NDCG@N considers not only whether items preferred by users are present in the top-N recommendation list but also accounts for their relative positions within that list, building upon the foundation of Recall@N.

$$NDCG@N = \sum_{u \in \mathcal{U}^{te}} \frac{NDCG_u@N}{|\mathcal{U}^{te}|}, \quad (22)$$

$$NDCG_u@N = \frac{DCG_u@N}{Z_u}, \quad (23)$$

where $DCG_u@N = \sum_{\ell=1}^{k} \frac{2\delta(i(\ell) \in \mathcal{I}_u^{te}) - 1}{\log(\ell+1)}$, and $Z_U$ is the optimal condition for $DCG_u@N$ in the ideal scenario.

### 4.3 Baselines

We will compare IAFMC with the following three base- line groups. The first group consists of static recommen- dation methods that ignore any sequence, i.e. BPRMF [29] and LightGCN [31]. The second group includes dynamic recommendation methods, i.e. FPMC [5] and Fossil [7]. The third group comprises deep learning algorithms, i.e. Caser and GRU4Rec [14].

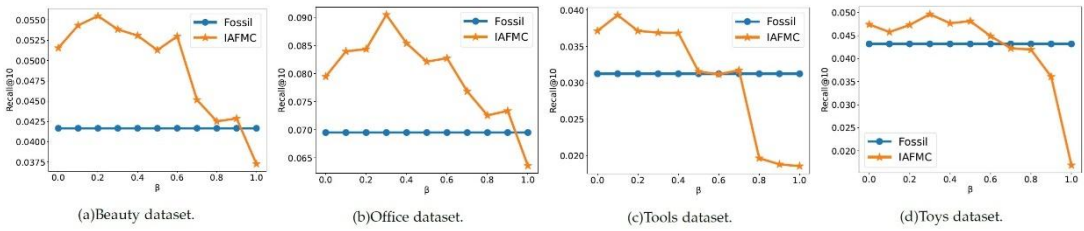

Fig. 2: Recommendation performance (Recall@10) over various $\beta$ on all datasets.

### 4.4 Parameter Settings

In order to make a fair comparison, we fix the embedding dimension d=20, learning rate $\lambda = \{0.01, 0.001\}$, and the number of iterations $T = \{100, 300, 500\}$ as the basic parameters for all models. We will search for the optimal $L_2$ regularization parameter from {0.1, 0.01, 0.001}. Specific parameter settings for different models are as follows:

- **IAFMC**: For the search of the value of hyperparameter $\beta$ in range [0, 1], he optimal sequence length will be searched from $L = \{1, 2, 3\}$.

- **BPRMF**: BPRMF is the most classic implicit feedbackpersonalized ranking collaborative filtering method without the need for additional parameter settings.

- **LightGCN**: One of the present most advanced static recommendation methods currently, it considers high-order collaborative signals in the user-item graph, and we set the number of layers as {1, 2, 3}.

- **FPMC**: One of the most classic dynamic algorithms, it considers low-order Markov chain information without the need for additional parameter settings.
- **Fossil**: By considering short-term sequence information through high-order Markov chains, we will also search for the optimal sequence length from $L = \{1, 2, 3\}$.
- **Caser**: A CNN-based sequential recommendation method, searching for a length of sequence of $\{5, 10\}$.
- **GRU4Rec**: A sequential recommendation method based on DNN, searching for hidden size $=\{50, 100\}$

## 4.5 Results

### 4.5.1 Number of Trained parameters

The training parameter count for Fossil is $2md + m + (1 + n)L$, while for IAFMC it is $2md + m + (1 + n)L + m + 1$. According to the data in Table 2, it can be calculated that in the various datasets used in the experiment, the increase in training parameter count for IAFMC relative to Fossil by 2.12% to 2.17%.

### 4.5.1 Effect of The Hyperparameter β

In order to identify the impact of different parameters in the model on the results, we use Recall@10 to evaluate the performance of the data and select appropriate parameters. The hyperparameter $\beta \in [0,1]$. Then, we conduct three tests for different parameters and obtain the average results.

According to Fig. 2, we can see that when $β$ takes certain values, our IAFMC outperforms Fossil in terms of NDCG@5 performance on the test data. By adjusting different values of $β$, we can see that the best performance is achieved when $β$ is 0.2(Beauty), 0.3(Office), 0.1(Tools), 0.3(Toys). In this paper, we employ the value of $β$ to be in the range of [0.1, 0.3].

### 4.5.2 Overall Results

We compared the performance of all models in Table 3 and demonstrated the effectiveness of IAFMC. We explain the results through the following observations.

- IAFMC achieved the best performance on almost all metrics across all datasets, except for NDCG@5 in the Toys dataset. The improvement in most metrics ranged from 4.48% to 19.44%, demonstrating the effectiveness of IAFMC. Particularly, there was a notable improvement in Recall@10, with enhancements ranging from 7.18% to 19.44%. This improvement is attributed to the model's ability to extract and utilize the correlation information between items, enabling better discovery of item relevance. The model identifies item categories that the target user preferences and recommends similar items to the target user, leading to better performance in longer unranked recommendation lists, i.e. a characteristic well-reflected by the Recall@10 metric. Conversely, IAFMC's performance in the NDCG@5 metric, especially in the Toys dataset, was relatively average.

- BPRMF is a relatively simple static collaborative filtering algorithm, but its performance is poor because it does not consider time series information. On the other hand, FPMC achieves good results in various metrics on datasets other than Tools as it considers time series information. In the Tools dataset, Light- GCN is better at predicting user preferences, while Fossil performs better in ranking preferred items. The neural network algorithms Caser based on CNN and GRU4Rec based on DNN have relatively average performance.

- We also studied the impact of different number of iterations on model performance under the same learning rate. In Fig. 3, we can see that for the dataset used in the experiment, IAFMC is in a period of rapid learning of data features when the number of iterations is between 0 and 100. During this period, performance rapidly improves with an increase in the number of iterations, thanks to the refinement of the relationship information between items before training the model to accelerate the learning process. After 100 iterations, IAFMC enters a slow learning phase, and performance fluctuates as the number of iterations increases. Compared to the base line Fossil model, IAFMC has the advantage of faster learning and better performance.

TABLE 3: Overall Performance Comparison Table. The best results of the baseline are underlined. 'Improve.' is the relative improvement against the best baseline performance.

| Dataset | Metric | BPRMF | LightGCN | FPMC | Fossil | Caser | GRU4Rec | IAFMC | Improv. |
|---|---|---|---|---|---|---|---|---|---|
| Beauty | Recall@5 | 0.0162 | 0.0259 | 0.0339 | 0.0278 | 0.0095 | 0.0300 | 0.0385 | 13.56% |
| | NDCG@5 | 0.0101 | 0.0162 | 0.0227 | 0.0176 | 0.0062 | 0.0214 | 0.0245 | 7.92% |
| | Recall@10 | 0.0276 | 0.0464 | 0.0507 | 0.0445 | 0.0177 | 0.0423 | 0.0590 | 16.37% |
| | NDCG@10 | 0.0137 | 0.0228 | 0.0281 | 0.0230 | 0.0088 | 0.0253 | 0.0311 | 10.67% |
| Office | Recall@5 | 0.0171 | 0.0269 | 0.0475 | 0.0399 | 0.0328 | 0.0424 | 0.0523 | 10.10% |
| | NDCG@5 | 0.0117 | 0.0189 | 0.0309 | 0.0255 | 0.0218 | 0.0283 | 0.0347 | 12.29% |
| | Recall@10 | 0.0240 | 0.0405 | 0.0738 | 0.0623 | 0.0557 | 0.0652 | 0.0791 | 7.18% |
| | NDCG@10 | 0.0139 | 0.0232 | 0.0394 | 0.0327 | 0.0290 | 0.0357 | 0.0433 | 9.89% |
| Tools | Recall@5 | 0.0112 | 0.0210 | 0.0201 | 0.0204 | 0.0120 | 0.0174 | 0.0247 | 17.61% |
| | NDCG@5 | 0.0073 | 0.0139 | 0.0139 | 0.0141 | 0.0088 | 0.0122 | 0.0163 | 15.60% |
| | Recall@10 | 0.0180 | 0.0324 | 0.0310 | 0.0319 | 0.0236 | 0.0233 | 0.0387 | 19.44% |
| | NDCG@10 | 0.0094 | 0.0175 | 0.0174 | 0.0178 | 0.0125 | 0.0141 | 0.0208 | 16.85% |
| Toys | Recall@5 | 0.0147 | 0.0283 | 0.0345 | 0.0260 | 0.0145 | 0.0296 | 0.0367 | 6.37% |
| | NDCG@5 | 0.0098 | 0.0188 | 0.0245 | 0.0172 | 0.0093 | 0.0222 | 0.0240 | -0.02% |
| | Recall@10 | 0.0236 | 0.0457 | 0.0485 | 0.0399 | 0.0254 | 0.0367 | 0.0563 | 16.08% |
| | NDCG@10 | 0.0126 | 0.0244 | 0.0290 | 0.0217 | 0.0128 | 0.0245 | 0.0303 | 4.48% |

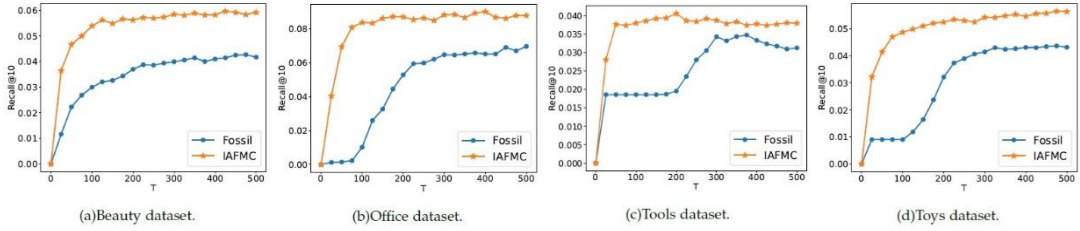

Fig. 3: Recommendation performance (Recall@10) with different iteration number $T$ on all datasets.

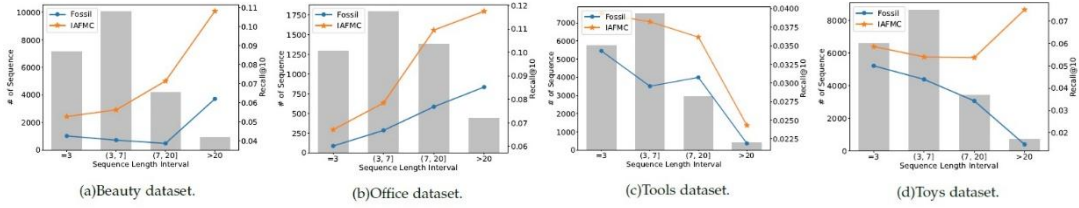

Fig. 4: Recommendation performance (Recall@10) on different sequence lengths of users on all datasets.

### 4.5.1 Improvements Analysis

We analyze the source of performance improvement by comparing the IAFMC of different user groups and projects. The analysis confirms the effectiveness of incorporating item association information in modeling and mitigating the cold start problem.

1) *Performances w.r.t Different Sequence Lengths of users.* We divided users into different groups based on their interaction frequency with items in the training section, i.e., the length of the user's training sequence. We will display the average Recall@10 for each group of users. Fig. 4 shows the size of each user group and the corresponding Recall@10 performance. We can see that the majority of users have interacted with items fewer than or equal to 7 times, with very few users having more than 20 interaction records.

From Figure 4, it can be seen that compared to Fossil, IAFMC shows the most significant improvement for users with interaction sequence lengths greater than 20. Across all datasets, the improvement for users ranges from 11.11% to 409.09%. We postulate the reason for these improvements that users with longer interaction sequences may have a wider variety of interests, and IAFMC is better able to capture the associations between these diverse interests and the corresponding items.

*2) Performances w.r.t Different interaction times of items.*

Here, we studied the performance of items based on differ- ent interaction frequencies, which indicates the popularity of different items as shown in Table 4. Most items in the validation set have been interacted with very few times. Among the items with 3 or fewer interactions, the Recall@10 value for Fossil is almost 0, while it ranges from 7‰ to 81‰ for IAFMC. Items with very few interactions are more strongly associated with some frequently interacted items, which lead to multiple training instances in IAFMC allowing our model to better model items with fewer interactions. The good performance of IAFMC on items with 3 or fewer interactions also demonstrates the model's advantage in addressing the cold start problem.

## 5 CONCLUSION

The paper proposes a new algorithm, Item Association Factorization Mixed Markov Chains for Sequential Recommendation for personalized recommendation. This algorithm introduces item association graph information based on Fossil to better extract the similarity and the item associations. The advantage of this method is that it achieves better recommendation performance with fewer additional parameters to cap the computational times. Our experiments show that this method performs well on four real datasets and effectively mitigates the cold start problem, demonstrating the importance and practicality of incorporating item association graph information. For future work, we will plan to integrate higher-order item association in- formation into the model and also integrate item association graph information into deep learning-based methods.